\documentclass{article}
\usepackage{latexsym}

\usepackage{amssymb}

\newcommand{\R}{\mathbb{R}}

\newcommand{\T}{\mathbb{T}}
\newcommand{\f}{f_{\infty}}

\def\prfe{\hspace*{\fill} $\Box$

\smallskip \noindent}
\newtheorem{Theorem}{Theorem}
\newtheorem{Proposition}{Proposition}
\newtheorem{Lemma}{Lemma}
 
\newtheorem{Remark}{Remark}

\title{The Newtonian limit of the relativistic \\ Boltzmann equation}
\author{Simone Calogero\\[0.5cm]
Max Planck Institut f\"ur Gravitationphysik\\
Albert Einstein Institut Am M\"uhlenberg 1\\
14476 Golm bei Potsdam, Germany\\
E-mail: simcal1@aei-potsdam.mpg.de}
\date { }
                                                                              
\begin{document} 
\maketitle
\begin{abstract} The relativistic Boltzmann equation for a constant differential cross section and with periodic boundary conditions 
is considered. The speed of light
appears as a parameter $c>c_0$ for a properly large and positive $c_0$. A local existence and uniqueness theorem is proved in an interval of time
independent of $c>c_0$ and conditions are given such that in the limit $c\to +\infty$ the solutions converge, in a suitable norm, to the
solutions of the non-relativistic Boltzmann equation for hard spheres.    
\end{abstract}

\section{Introduction}\label{intro}
The purpose of this paper is to show that solutions of the relativistic
Boltzmann equation are well-approximated by solutions of the
classical (non-relativistic) Boltzmann equation. A more precise statement will be given
later in the introduction.

The relativistic Boltzmann equation can be written in the form
\begin{equation}\label{relbol}
\partial_t f+\widehat{p}\cdot\nabla_x f= Q_\mathrm{rel}(f,f),
\end{equation}
where the various symbols have the following meaning. $f=f(t,x,p)$ is the distribution function in phase-space of a single non-degenerate 
relativistic gas. $\widehat{p}=cp/p_0$ is the relativistic velocity, with $c$ denoting the speed of
light and $y_0=\sqrt{c^2+|y|^2}$. The molecular rest-mass is set to unity and the convention for the signature of Minkowski's metric is $(+---)$, so that $p_0=p^0$. Finally,
$Q_\mathrm{rel}$ is the relativistic collision operator defined by
\begin{equation}\label{relcolopera}
Q_{\rm rel}(f,g)=\int_{{\R}^{3}}\int_{{S}^{2}} \mathcal{K}_c(p,q,\omega)[f(p') g(q')-f(p)g(q)]\,{\rm d}\omega{\rm d}q. 
\end{equation}
In the previous definition, $p',q'$ are the momenta after the elastic collision of two particles with pre-collisional momenta $p,q$. These quantities are
subjected to the conservation of momentum and energy, which read
\begin{equation}\label{cons}
p+q=p'+q',\quad \mathcal{E}_c(p)+\mathcal{E}_c(q)=\mathcal{E}_c(p')+\mathcal{E}_c(q'),\quad \mathcal{E}_c(y)=cy_0.
\end{equation}
A solution of (\ref{cons}) can be represented as
\begin{equation}\label{relgeom}
p'=p-a(p,q,\omega)\omega, \quad q'=q+a(p,q,\omega)\omega,
\end{equation}
where
\[
a(p,q,\omega)=\frac{2(p_0+q_0)[c^{-1}\omega\cdot(\widehat{p}-\widehat{q})]p_0q_0}{(p_0+q_0)^2-[\omega\cdot(p+q)]^2}.
\]
Using this representation, the relativistic collision kernel $\mathcal{K}_c$ takes the form
\begin{equation}\label{relcross}
\mathcal{K}_c(p,q,\omega)=16\sigma (c^2+\mathfrak{g}^2)\frac{(p_0+q_0)^2|\omega\cdot(\widehat{p}-\widehat{q})|}
{[(p_0+q_0)^2-(\omega\cdot(p+q))^2]^2},
\end{equation}
\begin{equation}\label{g}
\mathfrak{g}=\frac{1}{\sqrt{2}}(p_0q_0-p\cdot q-c^2)^{1/2}.
\end{equation} 
Here $\mathfrak{g}$ is a Lorentz invariant defined so that $-2\mathfrak{g}$ is the relative momentum in the center of mass system and $\sigma$ denotes the 
differential cross section. In general $\sigma$ is a function of $\mathfrak{g}$ and of a second Lorentz invariant quantity which in the center of mass system reduces to the cosine of the
scattering angle of the collision. In this article the differential cross section is assumed to be constant. 
As usual, the local dependence on $(t,x)$ in (\ref{relcolopera}) is omitted. This formulation of the relativistic Boltzmann equation for $c=1$ is
derived for instance in \cite{GS1,GS2}. We refer to \cite{A,C,GLW,S} for more background on the subject.

In this paper, the solutions of (\ref{relbol}) will be directly compared to the solutions of the classical Boltzmann equation for hard spheres, which is
\begin{equation}\label{clabol}
\partial_t \f+p\cdot\nabla_x \f= Q_\mathrm{cl}(\f,\f),
\end{equation}
where 
\begin{equation}\label{clacolopera}
Q_{\rm cl}(f,g)=d\int_{{\R}^{3}}\int_{{S}^{2}} |\omega\cdot(p-q)|[f(\bar{p}) g(\bar{q})-f(p)g(q)]\,{\rm d}\omega{\rm d}q. 
\end{equation}

The meaning of the various symbols in (\ref{clabol}) is the same as in the relativistic case. The post-collisional momenta are now denoted 
by $\bar{p},\bar{q}$, the conservation of momentum and energy take the form $p+q=\bar{p}+\bar{q}$ and $\frac{1}{2}|p|^2+\frac{1}{2}|q|^2=
\frac{1}{2}|\bar{p}|^2+\frac{1}{2}|\bar{q}|^2$
respectively, while the analogue of (\ref{relgeom}) in the classical case is 
\begin{equation}\label{clageom}
\bar{p}=p-\omega\cdot(p-q)\omega,\quad \bar{q}=q+\omega\cdot(p-q)\omega. 
\end{equation}
The factor $d$ in (\ref{clacolopera}) is the differential cross section for hard spheres interaction, which is a constant with the same dimensions as $\sigma$, 
namely $[length]^2$. A standard mathematical reference for the classical Boltzmann equation is \cite{CIP}.

In this paper it is shown that there is a class of solutions to the
relativistic Boltzmann equation which have a Newtonian limit, i.e., which tend to solutions of the classical Boltzmann equation (in the
corresponding class) as the speed of light goes to infinity. For this purpose, the speed of light will be treated as a parameter
$c>c_0$---where $c_0$ is a fixed and properly large positive constant--- and
the difference between the classical and the relativistic solution will be estimated as $c\to +\infty$. (No loss of generality arises in
letting $c>c_0$, since only the limit behaviour as $c\to+\infty$ is of interest here). 
In order to obtain the correct Newtonian limit it is also necessary to relate the constants $d$ and $\sigma$ in a proper way. 
From (\ref{relcross}) it follows that $\mathcal{K}_c(p,q,\omega)\to 4\sigma|\omega\cdot(p-q)|$ as $c\to +\infty$. 
This leads to {\it postulate} the relation $4\sigma = d$. By further choosing units such that $d=1$, 
the classical collision kernel reduces to $|\omega\cdot (p-q)|$, while the relativistic collision kernel becomes
\begin{equation}\label{relcros2}
\mathcal{K}_c(p,q,\omega)=2(p_0q_0-p\cdot q+c^2)\frac{(p_0+q_0)^2|\omega\cdot(\widehat{p}-\widehat{q})|}{[(p_0+q_0)^2-(\omega\cdot(p+q))^2]^2}.
\end{equation}

The precise formulation of the result will be now given. The conditions on the distribution functions mentioned in the 
theorem are introduced thereafter. The symbol $\T^3$ denotes the three-torus and the norm $\|\ \|_{0,1}$ is defined as follows:
\[
\|g(t)\|_{0,1}=\int_{\R^3}|g(t,p)|_0 {\rm d}p,\ \ |g(t,p)|_0=\sup_{x\in \T^3}|g(t,x,p)|.
\]
\begin{Theorem}\label{main}
Let $\f(t)$ be a solution of (\ref{clabol}) which satisfies the properties (C1),(C2) and with initial datum 
$\f^{\rm in}\in C^1$ such that $|\nabla_p\f^{\rm in}|_0\in L^1(\R^3)$.
Let $f_c(t)$ be a solution of (\ref{relbol}), depending on $c>c_0$, 
which satisfies the properties (R1)--(R3) and with the $c$-dependent initial datum 
$f^{\rm in}_c$. Assume $\|\f^{\rm in}- f^{\rm in}_c\|_{0,1}=O(c^{-1})$ as $c\to +\infty$. Then
\[
\lim_{c\to +\infty}\|\f(t)- f_c(t)\|_{0,1}= 0, \ \ t\in [0,T].
\]
\end{Theorem} 

The following notation will be used. Given two functions $g$ and $h$ on $\mathbb{R}^n$ we write $g\lesssim h$ if the 
estimate $g\leq D h$ holds for a positive constant $D$ independent of $c>c_0$.
The constant $D$ may also depend on the length of some time interval $[0, T]$, in which case we write
$g \lesssim h$ for $t \in [0, T]$. Whenever necessary or convenient, the constant $D$ will be recovered in the computations.

The classes of solutions of the Boltzmann equations to be considered are defined by the following properties.  In the classical case it
is required that, in some interval $[0,T_1]$,
\begin{itemize}
\item[(C1)] $\f\in C([0,T_1]\times\mathbb{T}^3\times\mathbb{R}^3)$,
\item[(C2)] $\exists\,\alpha_0>0: \f(t,x,p)\lesssim \exp(-\alpha_0 |p|^2), t\in [0,T_1], x\in\T^3, p\in\R^3$. 
\end{itemize}

In the relativistic case let $f_c$ denote a (one parameter family of) solution(s) of (\ref{relbol}) and require that, for all $c>c_0$ and in some interval $[0,T_c]$,
\begin{itemize}
\item[(R1)] $f_c\in C([0,T_c]\times\mathbb{T}^3\times\mathbb{R}^3)$,
\item[(R2)] $\exists\,\beta_0>0: f_c(t,x,p)\lesssim \exp\left[-\beta_0 (\mathcal{E}_c(p)-c^2)\right], t\in [0,T_c], x\in\T^3, p\in\R^3$,
\item[(R3)] $T_2:=\inf_{c>c_0} T_c>0$. 
\end{itemize}

Let us briefly comment on the above conditions.
The existence of solutions to the classical Boltzmann equation satisfying the properties (C1), (C2) is
proved in \cite{KS}---see also \cite{ACI, IS} for questions concerning the global existence of such solutions. A similar argument applies 
to the relativistic Boltzmann equation to prove the local existence and uniqueness of solutions
satisfying (R1)--(R3). A short sketch of the proof is given in section 3 to show that the property (R3) is satisfied. 
The latter is necessary for studying  the Newtonian limit, since it assures that the existence interval of a solution of the relativistic 
Boltzmann equation does not shrink to zero as $c\to +\infty$. The time $T$ in theorem \ref{main} is defined as the minimum between 
$T_1$ and $T_2$. 
 
Note also that the assumption $x\in \T^3$ allows one to neglect technical difficulties not related to the problem under discussion,
such us the choice of boundary or fall-off conditions. The generalization of the result when $x$ lies in a region of $\R^3$
with smooth boundary---or simply $x\in\R^3$---is not attempted here but it should not be too difficult. 

\section{Proof of the main theorem}
The following lemma collects some estimates which are required in the proof of the main theorem.
\begin{Lemma}\label{est}
The following estimates hold:
\begin{eqnarray*}
&&(a)\quad |q'-\bar{q}|+|p'-\bar{p}|\lesssim \frac{(|q|+|p|)^3}{c^2},\\
&&(b)\quad\big |\mathcal{K}_c(p,q,\omega)-|\omega\cdot(p-q)|\big|\lesssim \frac{(1+|p|+|q|)^{9}}{c^2},\\
&&(c)\quad\int\!\!\!\!\!\!\!\!\int\mathcal{K}_c(p,q,\omega)\exp{\left[-\beta_0(\mathcal{E}_c(q)-c^2)\right]}{\rm d}\omega{\rm d}q
\lesssim (1+|p|).
\end{eqnarray*}
\end{Lemma} 
\noindent\textit{Proof:} From (\ref{relgeom}) and (\ref{clageom}) we have 
\[
|q'-\bar{q}|=|p'-\bar{p}|=|\omega\cdot (p-q)-a|
\]
and a computation shows that
\[
\omega\cdot (p-q)-a=\frac{\omega\cdot (p+q)(|p|^2-|q|^2+(\omega\cdot p)^2-(\omega\cdot q)^2)}{(p_0+q_0)^2-[\omega\cdot(p+q)]^2}
=\frac{{\rm Num}}{{\rm Den}}.
\]
Moreover
\begin{eqnarray}
{\rm Den}&=&2c^2+|p|^2+|q|^2+2\sqrt{c^2+|p|^2}\sqrt{c^2+|q|^2}-(\omega\cdot p)^2-(\omega\cdot q)^2-2(\omega\cdot p)(\omega\cdot q)\nonumber\\
&\geq& 2c^2+2\sqrt{c^2+|p|^2}\sqrt{c^2+|q|^2}-2|p||q|\geq 2c^2+\frac{c^2(c^2+|p|^2+|q|^2)}{\sqrt{c^2+|p|^2}\sqrt{c^2+|q|^2}}\label{estD}.
\end{eqnarray}
In particular ${\rm Den}>2c^2$ and since ${\rm Num}\lesssim (|p|+|q|)^3$, the estimate (a) is proved. Next from (\ref{relcros2}) we have
\begin{eqnarray}
&&\big |\mathcal{K}_c(p,q,\omega)-|\omega\cdot(p-q)|\big|\lesssim
\left|\frac{2(p_0q_0-p\cdot q+c^2)(p_0+q_0)^2(\omega\cdot\widehat{p}-\omega\cdot\widehat{q})}{[(p_0+q_0)^2-(\omega\cdot(p+q))^2]^2}
-\omega\cdot(p-q)\right|\nonumber\\
&&\quad\quad\quad\lesssim |\omega\cdot p|\left|\frac{2c(p_0q_0-p\cdot
q+c^2)(p_0+q_0)^2}{p_0[(p_0+q_0)^2-(\omega\cdot(p+q))^2]^2}-1\right|+(q\leftrightarrow p),\label{temp}
\end{eqnarray}
where $(q\leftrightarrow p)$ denotes the expression obtained by exchanging $p$ and $q$ in the first term. Recall the definition of ``${\rm
Den}$" in (\ref{estD}). The first term in (\ref{temp}) is estimated as 
\begin{eqnarray*}
&&\frac{|p|}{p_0({\rm Den})^2}\Big|2c(p_0q_0+c^2)(p_0+q_0)^2-2cp\cdot q(p_0+q_0)^2-p_0(p_0+q_0)^4\\
&&\quad\quad -p_0[\omega\cdot(p+q)]^4+2p_0(p_0+q_0)^2[\omega\cdot(p+q)]^2\Big|\\
&&\lesssim\frac{|p|}{p_0({\rm Den})^2}\Big|(p_0+q_0)^2(2cp_0q_0+2c^3-p_0^3-p_0q_0^2-2p_0^2q_0)+c^3(1+|p|+|q|)^5\Big|.
\end{eqnarray*}
Here $p_0({\rm Den})^2\geq 4c^5$ and so to prove (b) one needs to estimate only the expression containing the fifth order powers of $c$, 
which is given by
\[
\mathcal{P}(p,q)=(p_0+q_0)^2[2cp_0q_0+2c^3-p_0^3-p_0q_0^2-2p_0^2q_0].
\]
We have
\[
\frac{|\mathcal{P}(p,q)|}{p_0({\rm Den})^2}\lesssim \frac{(1+|p|+|q|)^2}{c^3}\big|2cp_0q_0+2c^3-p_0^3-p_0q_0^2-2p_0^2q_0\big|.
\]
Using that
\[
\quad2cp_0q_0-2p_0^2q_0=\frac{2p_0q_0}{c+p_0}(c^2-p_0^2)\leq 2c|p|^2(1+|q|),
\]
\begin{eqnarray*}
c^3-p_0q_0^2&=&c^2(c-p_0)-c|q|^2(1+|p|)\\
&\leq&\frac{c^2(c^2-p_0^2)}{c+p_0}+c|q|^2(1+|p|)\leq c(|p|^2+|q|^2)(1+|p|),
\end{eqnarray*}
\[
c^3-p_0^3=c^3\left(1-\left(1+\frac{|p|^2}{c^2}\right)^{3/2}\right)\leq c(1+|p|)^6,
\]
and repeating the argument for the second term in (\ref{temp}) concludes the proof of (b). To prove (c) consider the following pointwise estimate on $\mathcal{K}_c$:
\begin{eqnarray*}
\mathcal{K}_c&\lesssim&\frac{cp_0q_0}{({\rm Den})^2}(p_0^2+q_0^2+2p_0q_0)(|p|/p_0+|q|/q_0)\\
&\lesssim&\frac{c}{({\rm Den})^2}[|p|p_0^2q_0+|p|q_0^3+2|p|p_0q_0^2+p_0^3|q|+p_0|q|q_0^2+2p_0^2q_0|q|]\\
&\lesssim&\frac{c^4}{({\rm Den})^2}(1+|p|)(1+|q|^2)^{3/2}\left(1+\frac{|p|^2}{c^2}\right).
\end{eqnarray*}
From (\ref{estD}) it follows that 
\[
{\rm Den}\geq c^2\frac{\sqrt{1+|p|^2/c^2}}{\sqrt{1+|q|^2/c^2}}.
\] 
Moreover, since $\sqrt{c^4+c^2|q|^2}-c^2\geq \frac{1}{2}(\sqrt{1+|q|^2}-1)$, then
\begin{equation}\label{crucial} 
\exp[-\beta_0(\mathcal{E}_c(q)-c^2)]\lesssim\exp[-\frac{\beta_0}{2}\sqrt{1+|q|^2}] 
\end{equation}
and so
\[
\int\!\!\!\!\!\!\!\!\int \mathcal{K}_c(p,q,\omega)e^{-\beta_0(\mathcal{E}_c(q)-c^2)}{\rm d}\omega{\rm d}q\lesssim (1+|p|)
\int (1+|q|^2)^{5/2}e^{-\frac{\beta_0}{2}\sqrt{1+|q|^2}} {\rm d}q,
\]
by which the claim follows.\prfe

\begin{Remark} \textnormal{The simple estimate (\ref{crucial}) will be often used in the sequel for the same purpose as in lemma 1, i.e., 
to obtain an estimate independent of $c>c_0$ of the integrals containing the factor $\exp{\left[-\beta_0(\mathcal{E}_c(q)-c^2)\right]}$}.
\end{Remark}
   
In the class of solutions that we are considering, the distribution functions satisfy the Boltzmann equations in the mild form:
\begin{equation}\label{mildrel}
f_c(t,x,p)=f^{\rm in}_c(x-\widehat{p}t,p)+\int_0^t Q_{\rm rel}(s,x+\widehat{p}(s-t),p){\rm d}s,
\end{equation}
\begin{equation}\label{mildcla}
\f(t,x,p)=\f^{\rm in}(x-pt,p)+\int_0^t Q_{\rm cl}(s,x+p(s-t),p){\rm d}s.
\end{equation}
We use this representation to estimate the following quantity:
\[
F_\eta[\f]=\int \sup_{|h|<\eta}|\f(p+h)-\f(p)|_0 {\rm d}p,\quad \eta>0.
\]
\begin{Lemma}\label{continuity}
For all $\eta_0,T>0$, $\eta\in [0,\eta_0]$ and $t\in [0,T]$, there exists a
positive constant $C=C(T,\eta_0,\alpha_0)$ such that
\[
F_\eta[\f]\leq C\sqrt{F_\eta[\f^{\rm in}]}.
\]
\end{Lemma}
\noindent\textit{Proof:} By (\ref{clacolopera}),
\[
Q_{\rm cl}(\f,\f)(p+h)=\int\!\!\!\!\!\!\!\!\int|\omega\cdot(p-q)|[\f(\bar{p}+h)\f(\bar{q}+h)-\f(p+h)\f(q+h)]{\rm d}\omega{\rm
d}q.
\]
Therefore by (\ref{mildcla}),
\begin{eqnarray*}
\lefteqn{|\f(p+h)-\f(p)|_0\leq |\f^{\rm in}(p+h)-\f^{\rm in}(p)|_0}\\
&&+\int_0^t\int\!\!\!\!\!\!\!\!\int|\omega\cdot(p-q)||\f(q)|_0|\f(p+h)-\f(p)|_0{\rm d}\omega{\rm d}q{\rm d}s\\
&&+\int_0^t\int\!\!\!\!\!\!\!\!\int|\omega\cdot(p-q)||\f(\bar{q})|_0|\f(\bar{p}+h)-\f(\bar{p})|_0{\rm d}\omega{\rm d}q{\rm d}s\\
&&+\int_0^t\int\!\!\!\!\!\!\!\!\int|\omega\cdot(p-q)||\f(p+h)|_0|\f(q+h)-\f(q)|_0{\rm d}\omega{\rm d}q{\rm d}s\\
&&+\int_0^t\int\!\!\!\!\!\!\!\!\int|\omega\cdot(p-q)||\f(\bar{p}+h)|_0|\f(\bar{q}+h)-\f(\bar{q})|_0{\rm d}\omega{\rm d}q{\rm d}s.
\end{eqnarray*}
Hence changing to the post-collisional variables,
\begin{eqnarray*}
F_\eta[\f]&\lesssim& F_\eta[\f^{\rm in}]+\int_0^t\int\!\!\!\!\!\!\!\!\int\!\!\!\!\!\!\!\!\int|\omega\cdot(p-q)|e^{-\alpha_0|q|^2}\sup_{|h|<\eta}|\f(p+h)-\f(p)|_0{\rm d}\omega{\rm d}q{\rm d}p{\rm d}s\\
&&+\int_0^t\sup_{|h|<\eta}\int\!\!\!\!\!\!\!\!\int\!\!\!\!\!\!\!\!\int|\omega\cdot(p-h-q)|e^{-\alpha_0|p|^2}|\f(q+h)-\f(q)|_0{\rm d}\omega{\rm d}q{\rm d}p{\rm d}s\\
&=&F_\eta[\f^{\rm in}]+{\rm A}+{\rm B}.
\end{eqnarray*}
For $R>0$ we write
\begin{eqnarray*}
{\rm A}&\lesssim&\int_0^t\int_{|p|\leq R}(1+|p|)\sup_{|h|<\eta}|\f(p+h)-\f(p)|_0 {\rm d}p {\rm d}s\\
&&+\int_0^t\int_{|p|>R}(1+|p|) \exp[-\alpha_0(|p|^2-2\eta|p|)]{\rm d}p {\rm d}s\\
&\lesssim& (1+R)\int_0^tF_\eta[\f](s){\rm d}s+Ce^{-\alpha_0 R^2/2}.
\end{eqnarray*}
The estimate for ${\rm B}$ is obtained in the same way,  
\[
{\rm B}\lesssim \left [1+(R+\eta_0)\right]\int_0^t F_\eta[\f](s) {\rm d}s +Ce^{-\alpha_0R^2/2}.
\]
Hence finally,
\[
F_\eta[\f](t)\lesssim F_\eta[\f^{\rm in}]+Ce^{-\alpha_0R^2/2}+\left [1+(R+\eta_0) \right]\int_0^t F_\eta[\f](s) {\rm d}s.
\]
Choose $R$ such that $e^{-\alpha_0R^2/2}=(1+F_{\eta_0}[\f^{\rm in}])^{-1} F_{\eta}[\f^{\rm in}]$, so that
\[
F_\eta[\f](t)\lesssim Ce^{-\alpha_0R^2/2}+\left [1+(R+\eta_0) \right]\int_0^tF_\eta[\f](s){\rm d}s.
\]
Hence, by the Gr\"onwall Lemma,
\begin{eqnarray*}
F_\eta[\f](t)&\lesssim& C\exp \left[-\frac{\alpha_0R^2}{2}+(R+\eta_0) t\right]\\
&\lesssim&
C\exp\left(-\frac{\alpha_0R^2}{4}\right)\sup_{R>0}\exp\left[-\frac{\alpha_0R^2}{4}+(R+\eta_0)T\right]\\
&\lesssim& C\exp\left(-\frac{\alpha_0R^2}{4}\right)=C\sqrt{F_\eta[\f^{\rm in}]}.
\end{eqnarray*}\prfe

Note also that for an initial datum as given in theorem \ref{main} the estimate $F_\eta[\f^{\rm in}]\lesssim\eta$ holds. Then lemma
\ref{continuity} implies
\begin{equation}\label{continuity2}
F_\eta[\f]\lesssim \sqrt{\eta}.
\end{equation} 

The next goal is to estimate the difference $Q_{\rm rel}-Q_{\rm cl}$ in the norm $\|\ \ \|_{0,1}$. 
\begin{Lemma}\label{estQ}
The following estimate holds:
\begin{eqnarray*}
\|Q_{\rm rel}(t)-Q_{\rm cl}(t)\|_{0,1}&\lesssim& c^{-1}(\log c)^{5/4}+\exp\left[-\beta_0(\sqrt{c^4+c^2\log c}-c^2)\right]\\
&&+\exp\left[-\alpha_0\log
c\right]+\sqrt{\log c}\|\f(t)-f_c(t)\|_{0,1}.
\end{eqnarray*}
\end{Lemma}
\noindent\textit{Proof:} From (\ref{relcolopera}) and (\ref{clacolopera}),
\begin{eqnarray*}
\|Q_{\rm rel}(t)-Q_{\rm cl}(t)\|_{0,1}&\lesssim& \int\!\!\!\!\!\!\!\!\int\!\!\!\!\!\!\!\!\int\Big |\mathcal{K}_c(p,q,\omega)[f(p')f(q')-f(p)f(q)]\\
&&-|\omega\cdot(p-q)|[\f(\bar{p})\f(\bar{q})-\f(p)\f(q)]\Big|_0{\rm d}\omega{\rm d}q{\rm d}p\\
&=& \int\!\!\!\!\!\!\!\!\int\!\!\!\!\!\!\!\!\int_{|p|+|q|\leq \sqrt{\log c}}\cdots+\int\!\!\!\!\!\!\!\!\int\!\!\!\!\!\!\!\!\int_{|p|+|q|>\sqrt{\log c}}\cdots
\end{eqnarray*}
Observing conservation of energy and using (c) of lemma \ref{est}, the integral in the exterior region is dominated by
\begin{eqnarray*}
\lefteqn{\int\!\!\!\!\!\!\!\!\int\!\!\!\!\!\!\!\!\int_{|p|+|q|>\sqrt{\log c}}\mathcal{K}_c(p,q,\omega)
\exp\left[-\beta_0(\mathcal{E}_c(p)+\mathcal{E}_c(q)-2c^2)\right]{\rm d}\omega{\rm d}q{\rm d}p}\\
&&+\int\!\!\!\!\!\!\!\!\int\!\!\!\!\!\!\!\!\int_{|p|+|q|>\sqrt{\log c}}|\omega\cdot(p-q)|\exp[-\alpha_0(|p|^2+|q|^2){\rm d}p{\rm d}\omega{\rm d}q{\rm d}p\\
&&\lesssim \exp\left[-\beta_0(\sqrt{c^4+c^2\log c}-c^2)\right]+\exp(-\alpha_0\log c).
\end{eqnarray*}
For the integral over the interior part consider the splitting
\[
\int\!\!\!\!\!\!\!\!\int\!\!\!\!\!\!\!\!\int_{|p|+|q|\leq \sqrt{\log c}}\cdots\leq {\rm I}+{\rm II}+...{\rm VIII},
\] 
where
\begin{eqnarray*}
&&{\rm I}=\int\!\!\!\!\!\!\!\!\int\!\!\!\!\!\!\!\!\int_{|p|+|q|\leq \sqrt{\log
c}}|f_c(p)|_0|f_c(q)|_0\Big||\omega\cdot(p-q)|-\mathcal{K}_c\Big|{\rm d}\omega{\rm d}q{\rm d}p,\\
&&{\rm II}=\int\!\!\!\!\!\!\!\!\int\!\!\!\!\!\!\!\!\int_{|p|+|q|\leq \sqrt{\log
c}}|\f(\bar{p})|_0|\f(\bar{q})|_0\Big||\omega\cdot(p-q)|-\mathcal{K}_c\Big|{\rm d}\omega{\rm d}q{\rm d}p,\\
&&{\rm III}=\int\!\!\!\!\!\!\!\!\int\!\!\!\!\!\!\!\!\int_{|p|+|q|\leq \sqrt{\log c}} |\omega\cdot(p-q)||\f(p)|_0|\f(q)-f_c(q)|_0{\rm d}\omega{\rm d}q{\rm d}p,\\
&&{\rm IV}=\int\!\!\!\!\!\!\!\!\int\!\!\!\!\!\!\!\!\int_{|p|+|q|\leq \sqrt{\log c}} |\omega\cdot(p-q)||f_c(q)|_0|\f(p)-f_c(p)|_0{\rm d}\omega{\rm d}q{\rm d}p,\\
&&{\rm V}=\int\!\!\!\!\!\!\!\!\int\!\!\!\!\!\!\!\!\int_{|p|+|q|\leq \sqrt{\log c}} \mathcal{K}_c|f_c(p')|_0|f_c(q')-\f(q')|_0{\rm d}\omega{\rm d}q{\rm d}p,\\
&&{\rm VI}=\int\!\!\!\!\!\!\!\!\int\!\!\!\!\!\!\!\!\int_{|p|+|q|\leq \sqrt{\log c}} \mathcal{K}_c|f_c(p')|_0|\f(q')-\f(\bar{q})|_0{\rm d}\omega{\rm d}q{\rm d}p,\\
&&{\rm VII}=\int\!\!\!\!\!\!\!\!\int\!\!\!\!\!\!\!\!\int_{|p|+|q|\leq \sqrt{\log c}} \mathcal{K}_c|\f(\bar{q})|_0|f_c(p')-\f(p')|_0{\rm d}\omega{\rm d}q{\rm d}p,\\
&&{\rm VIII}=\int\!\!\!\!\!\!\!\!\int\!\!\!\!\!\!\!\!\int_{|p|+|q|\leq \sqrt{\log c}} \mathcal{K}_c|\f(\bar{q})|_0|\f(p')-\f(\bar{p})|_0{\rm d}\omega{\rm d}q{\rm d}p.
\end{eqnarray*}
It follows directly from the estimate (b) of lemma \ref{est} that 
\[
{\rm I}+{\rm II}\lesssim c^{-2}. 
\]
The integral ${\rm III}$ and ${\rm IV}$ satisfy the estimate
\[
{\rm III}+{\rm IV}\lesssim \int_{|q|\leq \sqrt{\log c}}(1+|q|)|\f(q)-f_c(q)|_0 {\rm d}q\lesssim (1+\sqrt{\log c})\|\f(t)-f_c(t)\|_{0,1}.
\]
In the integral ${\rm V}$ we change to the post-collisional variables. 
Since $\mathcal{K}_c(p,q,\omega){\rm d}q{\rm d}p=\mathcal{K}_c(p',q',\omega){\rm d}q'{\rm d}p'$ and, by (\ref{cons}), $|p'|+|q'|\leq 4\sqrt{\log c}$ for 
$|p|+|q|\leq \sqrt{\log c}$, then
\begin{eqnarray*}
{\rm V}&\lesssim& \int\!\!\!\!\!\!\!\!\int\!\!\!\!\!\!\!\!\int_{|p|+|q|\leq 4\sqrt{\log c}} \mathcal{K}_c(p,q,\omega) 
|f_c(p)|_0|f_c(q)-\f(q)|_0|{\rm d}\omega{\rm d}q{\rm d}p\\
&\lesssim& \sqrt{\log c}\|\f(t)-f(t)\|_{0,1}.
\end{eqnarray*}
For the integral ${\rm VI}$ we have, by the estimate (a) of lemma \ref{est}, lemma \ref{continuity} and (\ref{continuity2}),
\begin{eqnarray*}
{\rm VI}&\lesssim&\int\!\!\!\!\!\!\!\!\int\!\!\!\!\!\!\!\!\int_{|p|+|q|\leq \sqrt{\log c}}\mathcal{K}_c|f_c(p')|_0\sup_{|h|\lesssim
\frac{(\log c)^{3/2}}{c^2}}|\f(q'+h)-\f(q')|_0{\rm d}\omega{\rm
d}q{\rm d}p\\
&\lesssim& \int\!\!\!\!\!\!\!\!\int\!\!\!\!\!\!\!\!\int_{|p|+|q|\leq 4\sqrt{\log c}}\mathcal{K}_ce^{-\beta_0(\mathcal{E}(p)-c^2)}
\sup_{|h|\lesssim \frac{(\log c)^{3/2}}{c^2}}|\f(q+h)-\f(q)|_0{\rm d}\omega{\rm
d}q{\rm d}p\\
&\lesssim& \int_{|q|\leq 4\sqrt{\log c}}(1+|q|)\sup_{|h|\lesssim \frac{(\log c)^{3/2}}{c^2}}|\f(q+h)-\f(q)|_0\lesssim
\sqrt{\log c}\,F_{\frac{(\log c)^{3/2}}{c^2}}[\f]\\
&\lesssim& c^{-1}(\log c)^{5/4}.
\end{eqnarray*}
It is now straightforward to estimate ${\rm VII}$ and ${\rm VIII}$, therefore we merely state the result:
\[
{\rm VII}+{\rm VIII}\lesssim c^{-1}(\log c)^{5/4}+\sqrt{\log c}\|\f(t)-f(t)\|_{0,1}. 
\] 
Collecting the various bounds the claim follows.\prfe
The proof of theorem \ref{main} is now almost complete.
From (\ref{mildrel}) and (\ref{mildcla}) we have  
\[
\|f_c(t)-\f(t)\|_{0,1}\leq \|f^{\rm in}_c-\f^{\rm in}\|_{0,1}+\int_0^t\|Q_{\rm rel}(s)-Q_{\rm cl}(s)\|_{0,1}.
\]
Using lemma \ref{estQ} and applying Gr\"onwall's inequality one obtains, for $t\in [0,T]$
\begin{eqnarray*}
&&\|f_c(t)-\f(t)\|_{0,1}\lesssim\|f^{\rm in}_c-\f^{\rm in}\|_{0,1}e^{D\sqrt{\log c}}+c^{-1}(\log c)^{5/4}e^{D\sqrt{\log c}}\\
&&\quad+\exp\left[-\beta_0\sqrt{c^4+c^2\log c}+\beta_0c^2+D\sqrt{\log c}\right]+\exp\left[-\alpha_0\log c+D\sqrt{\log c}\right].
\end{eqnarray*}
The expression in the right hand side tends to zero as $c\to\infty$ and this concludes the proof of theorem \ref{main}.

\section{Existence in a uniform short time interval}
The equation to be studied reads explicitly
\begin{equation}\label{mild}
f_c(t,x,p)=f^{\rm in}_c(x-\widehat{p}t,p)+\int_0^t[Q^{+}_{\rm rel}(f_c,f_c)-Q^{-}_{\rm rel}(f_c,f_c)](s,x-\widehat{p}(t-s),p){\rm d}s,
\end{equation}
where $Q^{+}_{\rm rel}$ and $Q^{-}_{\rm rel}$ refer to the gain and loss part of the relativistic collision operator (\ref{relcolopera}),
respectively.
Given two functions $u_0(t)$ and $l_0(t)$, the approximation sequences $\{u_n\}_{n\geq 0}$, $\{l_n\}_{n\geq 0}$ are defined recursively by
\[
l_{n+1}(t,x,p)=f^{\rm in}_c(x-\widehat{p}t,p)+\int_0^t[Q^{+}_{\rm rel}(l_n,l_n)-Q^{-}_{\rm
rel}(l_{n+1},u_n)](s,x-\widehat{p}(t-s),p){\rm d}s,
\]
\[
u_{n+1}(t,x,p)=f^{\rm in}_c(x-\widehat{p}t,p)+\int_0^t[Q^{+}_{\rm rel}(u_n,u_n)-Q^{-}_{\rm rel}(u_{n+1},l_n)](s,x-\widehat{p}(t-s),p){\rm d}s
\]
and as in lemma 5.1 in \cite{KS} one can prove the following
\begin{Proposition}
Assume the {\it beginning condition} is satisfied:
\begin{equation}\label{bc}
0\leq l_0(t)\leq l_1(t)\leq u_1(t)\leq u_0(t);
\end{equation}
then $0\leq l_n(t)\leq l_{n+1}(t)\leq u_{n+1}(t)\leq u_n(t)$ for all $n\geq 0$. 
\end{Proposition}
Next assume that $u_0\lesssim \exp[-\beta_0(\mathcal{E}_c(p)-c^2)]$; it follows by the previous proposition that 
$u_n,l_n$ are also dominated by $\exp[-\beta_0(\mathcal{E}_c(p)-c^2)]$. Moreover $l_n(t)\uparrow l(t)$, $u_n(t)\downarrow u(t)$ and 
$|u(t)|, |l(t)|\lesssim \exp[-\beta_0(\mathcal{E}_c(p)-c^2)]$. All these preliminary facts are valid for any collision kernel. When the
latter is given by (\ref{relcros2}) one can also prove that i) $u(t)=l(t)$ and ii) the limit is a continuous solution of (\ref{mild}). The second
statement is an obvious consequence of the first one, so only the proof of i) will be given. By the dominated convergence theorem, $u(t),
l(t)$ satisfy
\[
l(t,x,p)=f^{\rm in}_c(x-\widehat{p}t,p)+\int_0^t[Q^{+}_{\rm rel}(l,l)-Q^{-}_{\rm rel}(l,u)](s,x-\widehat{p}(t-s),p){\rm d}s,
\]
\[
u(t,x,p)=f^{\rm in}_c(x-\widehat{p}t,p)+\int_0^t[Q^{+}_{\rm rel}(u,u)-Q^{-}_{\rm rel}(u,l)](s,x-\widehat{p}(t-s),p){\rm d}s.
\]
Estimating the difference $u(t)-l(t)$ in the norm $\|\ \|_{0,1}$ and using (c) of lemma \ref{est} we get
\begin{eqnarray*}
\|u(t)-l(t)\|_{0,1}&\lesssim& \int_0^t\int (1+|p|)|u(s,p)-l(s,p)|_0 {\rm d}p {\rm d}s\\
&\lesssim& \int_0^t\int_{|p|\leq R}(1+|p|)|u(s,p)-l(s,p)|_0 {\rm d}p {\rm d}s\\
&&+\int_0^t\int_{|p|>R}(1+|p|)\exp[-\beta_0(\sqrt{c^4+c^2|p|^2}-c^2)] {\rm d}p {\rm d}s\\
&\lesssim& (1+R)\int_0^t\|u(s)-l(s)\|_{0,1}+te^{-\frac{\beta_0}{2}(\sqrt{c^4+c^2R^2}-c^2)}.
\end{eqnarray*} 
Hence by Gr\"onwall inequality
\[
\|u(t)-l(t)\|_{0,1}\lesssim te^t\exp\left[DtR-\frac{\beta_0}{2}\sqrt{c^4+R^2c^2}+\frac{\beta_0}{2}c^2\right],
\]
for some constant $D$ independent of the speed of light. For $t\in [0,T]$ and $c>\sqrt{48}DT/\beta_0:=c_0$, this implies
\[
\|u(t)-l(t)\|_{0,1}\lesssim e^{(-DTR)},\quad\textnormal{for }R>\frac{12DT}{\beta_0}
\]
and so the claim $u(t)=l(t)$, for $t\in [0,T]$ and $c> c_0$ follows by letting $R\to +\infty$. 

It remains to show that the beginning condition (\ref{bc}) is attained in some interval $[0,T_c]$ satisfying the property (R3) (indeed it
will be shown that $T_c$ is independent of $c>1$). Following \cite{KS} we choose $l_0\equiv 0$ and $u_0(t)$ of the form
\[
u_0(t)=\omega(t)e^{-\beta(t)[\mathcal{E}_c(p)-c^2]},
\]  
where $\beta$ and $\omega$ are positive functions and $\beta(0)=\beta_0$. We also set $\omega_0=\omega(0)$. Then $u_1$ are $l_1$ are given
by 
\[
l_1(t,x,p)=f^{\rm in}_c(x-\widehat{p}t,p)\exp\left[
-\omega(t)\int\!\!\!\!\!\!\!\!\int\mathcal{K}_c(q,p,\omega)e^{-\beta(t)[\mathcal{E}_c(q)-c^2]}{\rm d}\omega{\rm d}q\right],
\]
\[
u_1(t,x,p)=f^{\rm in}_c(x-\widehat{p}t,p)+\int_0^t\omega(s)^2\int\!\!\!\!\!\!\!\!\int\mathcal{K}_c(q,p,\omega)
e^{-\beta(s)[\mathcal{E}_c(p)+\mathcal{E}_c(q)-2c^2]}{\rm d}\omega{\rm d}q{\rm d}s.
\]
Hence $0\leq l_1(t)\leq u_1(t)$. Moreover 
\[
u_1(0)-u_0(0)=f^{\rm in}_c-\omega_0e^{-\beta_0[\mathcal{E}_c(p)-c^2]}\lesssim 
e^{-\beta_0[\mathcal{E}_c(p)-c^2]}-\omega_0e^{-\beta_0[\mathcal{E}_c(p)-c^2]}\leq 0,
\]
for $\omega_0$ large enough and
\begin{eqnarray*}
\frac{d}{dt}[u_1(t,x+\widehat{p}t,p)-u_0(t,x+\widehat{p}t,p)]&\leq&\left[
D(1+|p|)\omega^2-\dot{\omega}+\omega\dot{\beta}(\sqrt{c^4+c^2|p|^2}-c^2)\right]\\
&&\quad\times \exp[-\beta(t)(\mathcal{E}_c(p)-c^2)],
\end{eqnarray*}
where an upper dot has been used to denote differentiation in time. Hence the proof of (\ref{bc}) is complete if one can choose
$\omega,\beta$ such that 
\begin{equation}\label{final}
D(1+|p|)\omega^2-\dot{\omega}+\omega\dot{\beta}(\sqrt{c^4+c^2|p|^2}-c^2)\leq 0.
\end{equation}
Let 
\[
\omega(t)=\frac{\omega_0}{1-3D\omega_0t},\quad \beta(t)=\beta_0+\frac{2}{3}\log(1-3D\omega_0 t),
\]
so that $\dot{\omega}=3D\omega^2$ and $\dot{\beta}=-2D\omega$. Here $t\in [0,T]$, where $T=(6D\omega_0)^{-1}(1-e^{-\frac{3}{2}\beta_0})$ so that
$\omega$ and $\beta$ are well-defined positive functions in $[0,T]$. In this way, the left hand side of (\ref{final}) is dominated by 
$-D\omega^2\leq 0$ and this concludes the proof of the following
\begin{Theorem}
Let $f^{\rm in}_c\in C(\mathbb{T}^3\times\mathbb{R}^3)$ such that $f^{\rm in}_c\lesssim\exp[-\beta_0(\mathcal{E}_c(p)-c^2)]$.
 There exist $c_0,T>0$ such that for all $c>c_0$, the relativistic Boltzmann equation, eq. (\ref{relbol}) with $\mathcal{K}_c(p,q,\omega)$
 given by
(\ref{relcross}) and initial datum $f^{\rm in}_c$, has a unique solution $f\in C([0,T]\times\mathbb{T}^3\times\mathbb{R}^3)$ which
also satisfies $f\lesssim\exp[-\beta_0(\mathcal{E}_c(p)-c^2)]$; in particular the class of solutions satisfying (R1)--(R3) is
not empty.    
\end{Theorem} 
\noindent{\bf Acknowledgment:}
The author acknowledges the kind hospitality (and support) of the department of aeronautics and astronautics (graduate school of
engineering) of the Kyoto university. Support from the European HYKE network 
(contract HPRN-CT-2002-00282) is also acknowledged.

\end{document}